\documentclass[aps,twocolumn,prl,superscriptaddress,floatfix,showpacs]{revtex4}
\usepackage{amsmath}
\usepackage{graphicx}
\usepackage{subfigure}
\usepackage{color}
\usepackage{epstopdf}

\def\lsim{\mathrel{\mathpalette\gl@align<}}
\def\gsim{\mathrel{\mathpalette\gl@align>}}
\def\gl@align#1#2{\lower.6ex\vbox
{\baselineskip\z@skip\lineskip\z@
\ialign{$\m@th#1\hfil##\hfil$\crcr#2\crcr\sim\crcr}}}

\makeatother
\newcommand\ba{\begin{eqnarray}}
\newcommand\ea{\end{eqnarray}}
\newcommand\be{\begin{equation}}
\newcommand\ee{\end{equation}}
\newcommand\bi{\bibitem}
\newcommand{\ct}{\cite}

\def\non{\nonumber}

\def\si{\sigma}

\begin{document}

\title{Quenches and dynamical phase transitions in a non-integrable quantum Ising model}

\author{Shraddha Sharma}

\affiliation{Department of Physics, Indian Institute of
Technology, Kanpur 208 016, India}

\author{Sei Suzuki}
\affiliation{Department of Liberal Arts, Saitama Medical
University, Moroyama, Saitama 350-0495, Japan}

\author{Amit Dutta}

\affiliation{Department of Physics, Indian Institute of
Technology, Kanpur 208 016, India}

\date{\today}

\begin{abstract}
 We study quenching dynamics of a one-dimensional transverse Ising chain with nearest neighbor antiferromagentic
interactions in the presence of a longitudinal field which renders the model non-integrable. The dynamics of the spin chain is studied following a slow (characterized
by a rate) or sudden quenches of the longitudinal field; the residual energy, as obtained numerically using a t-DMRG scheme,   is found to satisfy 
analytically predicted scaling relations in both the cases. However, analyzing the temporal evolution of the Loschmidt overlap, we find different
possibilities of the presence (or absence) of dynamical phase transitions (DPTs) 
%at different instants of time, 
manifested in the non-analyticities of the rate function. Even though the model
is non-integrable, there are {periodic} occurrences of DPTs  when the system
is slowly ramped across the quantum critical point (QCP) as opposed to  the ferromagnetic (FM) version of the model; this numerical
finding is qualitatively explained  by mapping the original model to  an effective integrable spin model which is appropriate for describing such  slow quenches. Furthermore, concerning the sudden quenches, our numerical results  show that  in some 
cases, DPTs can be present even when the spin chain is quenched within
the same phase or even  to the QCP while in some
other situations they  completely disappear even after quenching across
the  QCP.  These observations 
lead us to the conclusion that it is the change in the nature of the ground state that determines the presence  of DPTs following
a sudden quench.
\end{abstract}

\pacs{75.10.Jm, 05.70.Jk, 64.60.Ht}

\maketitle
%%%%%%%%%%%%%%%%%%%%%%%%%%%%%%%%%%%%%%%%%%%%%%%%%%%%%%%%%%%%%%%%%%%%%

Following the remarkable advancement of the experimental studies of ultracold atoms  trapped in optical lattices \ct{bloch08,lewenstein12},
there is a recent upsurge in the studies of non-equilibrium dynamics of closed quantum systems, in particular
from the viewpoint of quantum quenches across a quantum critical point (QCP)\ct{Sachdev,suzuki13}. The relaxation time of the quantum
system diverges at the QCP resulting in a non-adiabatic dynamics and proliferation of topological defects in the
final state reached after the quench.

 According to the Kibble-Zurek (KZ) scaling relation \ct{Kibble1976,Zurek1985}, generalized to quantum critical 
systems \ct{Zurek2005,Polkovnikov2005}, when a $d$-dimensional quantum system, initially prepared in its ground state, is driven across an isolated QCP, by changing a 
parameter of the Hamiltonian
in a linear fashion as $t/\tau$, the density of defect satisfies the KZ scaling
 $\tau^{-d\nu/(z\nu + 1)}$; here, $\nu$ and $z$ are {the} correlation length and the dynamical exponent associated with the
 QCP respectively \ct{dziarmaga2005,damski05,cherng06}. Subsequently several modifications of the scaling have been proposed \ct{Mukherjee2007,Sengupta2008,Dutta2010}. Similarly when the system is quenched to the gapless QCP,  the residual energy (the excess energy over the ground state of the final Hamiltonian) 
scales as $\tau^{-(d + z)\nu/(z\nu + 1)}$;  on the contrary, when quenched to the gapped phase, the residual energy {follows} a scaling relation identical to that of
 the defect density. Similar scaling relations for the residual energy and the defect density have also been derived using an adiabatic perturbation theory for a sudden 
quench of small magnitude \ct{degrandi10}. (See  review articles [\onlinecite{PolkovnikovRev,dziarmaga10,dutta15}]).
 
It is well established that the phase transition in a thermodynamic system is marked by the non-analyticities in the free-energy density whose information can be 
obtained by analyzing the {zeros} of the partition function in a complex temperature plane as proposed by Fisher \ct{fisher65}. These
{zeros} of the partition function coalesce into a line (or area \ct{saarloos84}) in complex temperature plane, crossing the real axis in the 
thermodynamic limit; these crossings mark the non-analyticities in the
free-energy density. A similar observation was made earlier by Lee-Yang
\ct{lee52}  {for} a complex magnetic plane. In a similar spirit,
{a recent work by Heyl {\it et al.} \ct{heyl13}}
introduced the notion of {\it dynamical phase transitions} (DPTs) in connection to quantum quenches
probing  the non-analyticities in the \textit{dynamical free energy} in the complex  time plane. The idea stems from the similarity 
between the canonical partition function
\be
Z(\beta)= \textup{Tr}~ e^{-\beta H},
\ee
 of an equilibrium system (where $\beta$ is the inverse temperature) and that of the overlap amplitude (the Loschmidt overlap (LO)) defined at an instant of time $t$ as
\be
G(t)=\langle\psi_0|e^{-iHt}|\psi_0\rangle,
\ee
where, in the above {equation,} $H$ is the final Hamiltonian of the system reached through a sudden quenching of parameters, while  $|\psi_0\rangle$ is the ground state of the initial Hamiltonian. {Generalizing  to the complex  time   ($z$) plane, one can define the  dynamical free energy,
$f(z) = -\ln G(z)$}; one then  looks  for the {zeros} of the $G(z)$,  known as Fisher zeros, and  can claim the occurrence of  DPTs ({at real times})
 when  {the  lines of} Fisher zeros cross the {imaginary axis}. { These DPTs are manifested in sharp non-analyticities  in the rate function ($I(t) = - \ln |G(t)|^2/N$) at those instants of time}. This usually happens when the system is quenched across the QCP \cite{comment1}.  
% The
%non-analyticities are reflected in the rate function at instants $t=t_n^{*}$ where $n$ is an integer. 
The initial observation by  Heyl {\it{et al.}}
\ct{heyl13} {for a transverse Ising chain} led to a series of works  for both integrable and  the non-integrable spin chains  
\ct{karrasch13,andraschko14,kriel14} where DPTs were observed for  sudden quenches across the QCP. Although, a later work \ct{vajna14} showed that DPTs can
occur even when the system is quenched within the same phase. {These studies have also been generalized to 
two-dimensions 
%(where Fisher zeros
%coalesce into areas in the thermodynamic limit)} 
\ct{vajna15,schmitt15} 
where topology  may play a non-trivial role \ct{vajna15}.}

 A pertinent question at this point is how does the DPT depend on the integrability of the model under {consideration} or the nature of driving (slow or sudden)? Is
quenching across a QCP  essential to observe this? In this {letter}, we shall address these issues in the context
of a specific non-integrable model.
{We note in passing that the quantity  $|G(t)|^2$ denotes the Loschmidt echo which has been
studied in recent years in the context of decoherence  both in equilibrium \ct{quan06,rossini07,sharma12,cucchietti07} and non-equilibrium situations \ct{venuti10,mukherjee12,nag12,dora13} and has been generalized to finite temperature \ct{zanardi07_echo}. The LO is
also connected to the work-statistics \ct{gambassi11} and the entropy generation following a quench  \ct{dorner12}.}

The model we consider here is a {one-dimensional} Ising
model with a nearest neighbor antiferromagnetic ({AFM}) interaction $J$
(scaled to unity in the subsequent discussion) subjected to a transverse field ($\Gamma$) as well as a longitudinal field ($h$). {It} is described by the Hamiltonian \ct{ovchinnikov03}
\be
H= \sum_i \si_i^z \si_{i+1}^z - \Gamma \sum_i \si_i^x - h \sum_i \si_i^z,
\label{eq_ham}
\ee
where {$\si_i$'s} are Pauli matrices. For $h =0$, {the model is integrable with {QCPs} at $\Gamma=\Gamma_c= \pm 1$}, while any non-zero value of $h$  renders the model  non-integrable. Furthermore,
since the AFM interaction and the field $h$ compete with each other,
there is a QPT from AFM ordered phase to the disordered phase at a
particular value of  $\Gamma_c(h)$ for a given value {of} $h$. As a result, one  finds a phase diagram in the $\Gamma-h$ plane (separating the ordered from the disordered paramagnetic phase, see the supplementary material (SM)) starting from the integrable QCP (at $\Gamma_c=1$, $h=0$) at one  end and terminating {at  first order transition points at $\Gamma=0, h=\pm 2$ on the $h$-axis. }

In this letter, we shall restrict our attention to the case when
$\Gamma$ is fixed to $\Gamma_c=1$ {so that the system
{is} at the QCP when $h=0$;} $h$
{is driven}
slowly (i.e., defined by a rate {$\tau^{-1}$}) or suddenly in the vicinity of the  QCP when  the system
is always initially prepared in its ground state.
% This enables us, as
%we shall show below, to map the model in Eq.~(\ref{eq_ham}) to an equivalent model that would enable us to explain phenomenologically the numerically obtained results. 
In the presence of
a small $h$, a gap ($\Delta E$) opens up in the energy spectrum and a perturbation theoretic calculation, valid for small $h$, yields
$\Delta E \sim h^{\nu_h z}=h^2$ \ct{ovchinnikov03}, where $\nu_h$ is the correlation length exponent associated with the relevant perturbation $h$ and $z$ is the dynamical exponent
  associated with the QCP at $\Gamma=1$. Noting that $z=1$, one concludes that the exponent $\nu_h=2$.
 We note that a similar
study was reported in reference [\onlinecite{pollmann10}] {for a ferromagnetic (FM) Ising chain in a skewed field
(having both $\Gamma$
and $h$); however, the universal behavior associated with the FM case is different.}
%the QCP%where the
%quenching dynamics of a  subjected to a skewed field  \textcolor{red}{(having both $\Gamma$
%and $h$) which was ramped  across the QCP};   in the AFM case is different from that of the FM case where one finds that $\nu =8/15$ \ct{pollmann10}.
% 

%  
\begin{figure}
%\begin{center}
 \includegraphics[width=3.2in]{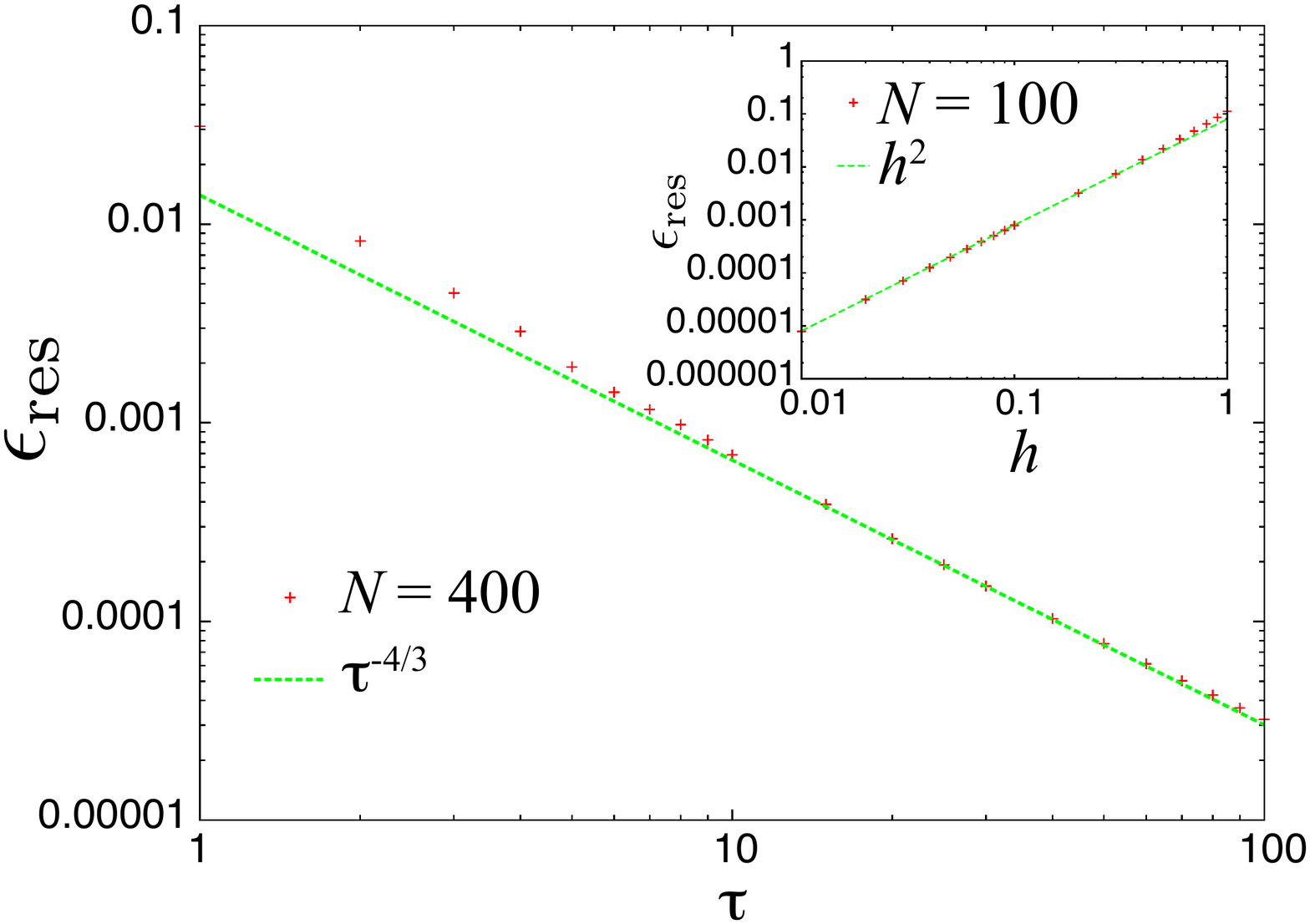}
% \end{center}
\caption{(Color online) Scaling of residual energy for slow and sudden quenching by changing the longitudinal field $h$ with the transverse 
        field $\Gamma=\Gamma_c=1$, as obtained from DMRG studies. When $h$ is changed linearly {as} $-t/\tau$ to the QCP ($h=0$), $\epsilon_{\rm res} \sim
        \tau^{-4/3}$ which is in perfect agreement with the KZ Scaling. In the Inset, we verify the scaling of $\epsilon_{\rm res} \sim h^2$ for a sudden quench starting from the QCP. }
     \label{fig:Eres}
\end{figure}

\begin{figure}
%\begin{center}
 \includegraphics[width=3.0in]{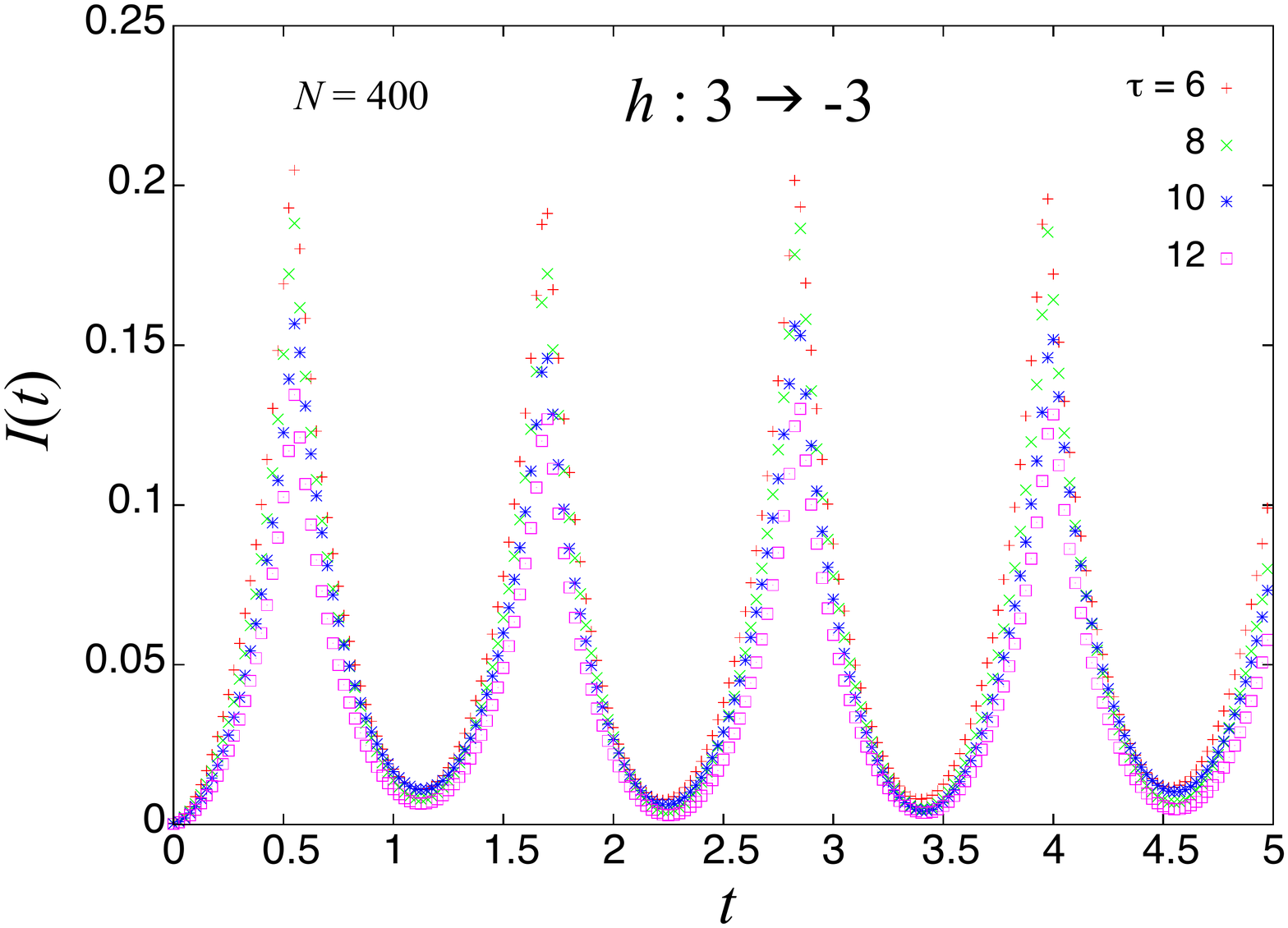}
% \end{center}
\caption{(Color online) Our numerical results show prominent periodic occurrences of DPTs when the longitudinal field is slowly ramped from a large positive value to
a large negative {value} as $ h\sim -t/\tau$. This periodic pattern can be qualitatively  explained  by studying $I(t)$ of the equivalent
integrable Hamiltonian (\ref{eq:ham_eqv}) as demonstrated in the SM. 
%\textcolor{red}{Sei, please provide
%with the data for quenching of $h$ from $1$ to $-1$; see supplementary material for justification, we shall push the 
%present figure to the supplementary material if required}.
}
     \label{fig:DPT2_tau}
\end{figure}

\begin{figure}[h!]
\centering
\subfigure[\  ]{
\includegraphics[width=4cm]{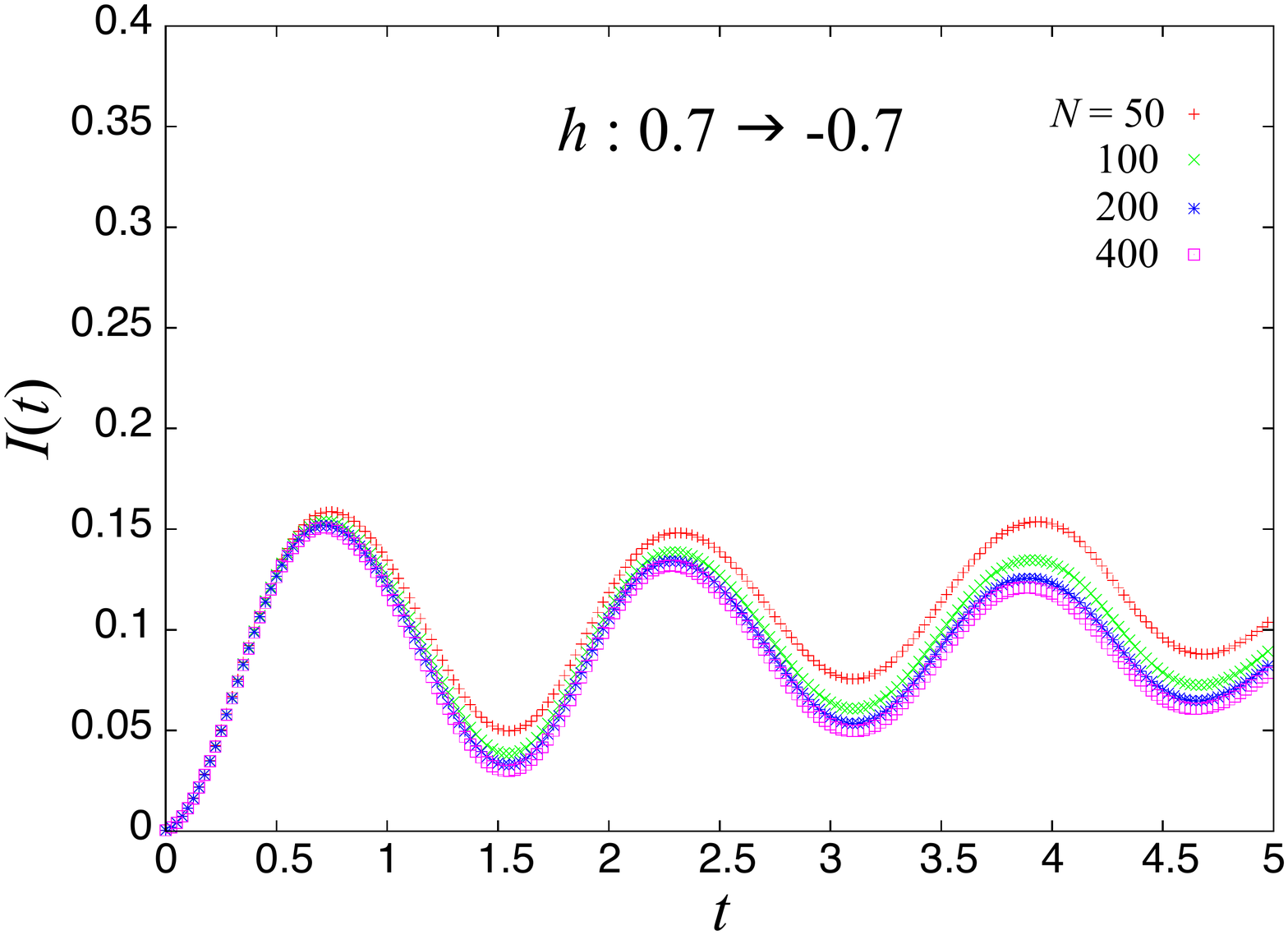}}
% \label{fig:DPT07_sud}
\subfigure[\ ]{
\includegraphics[width=4cm]{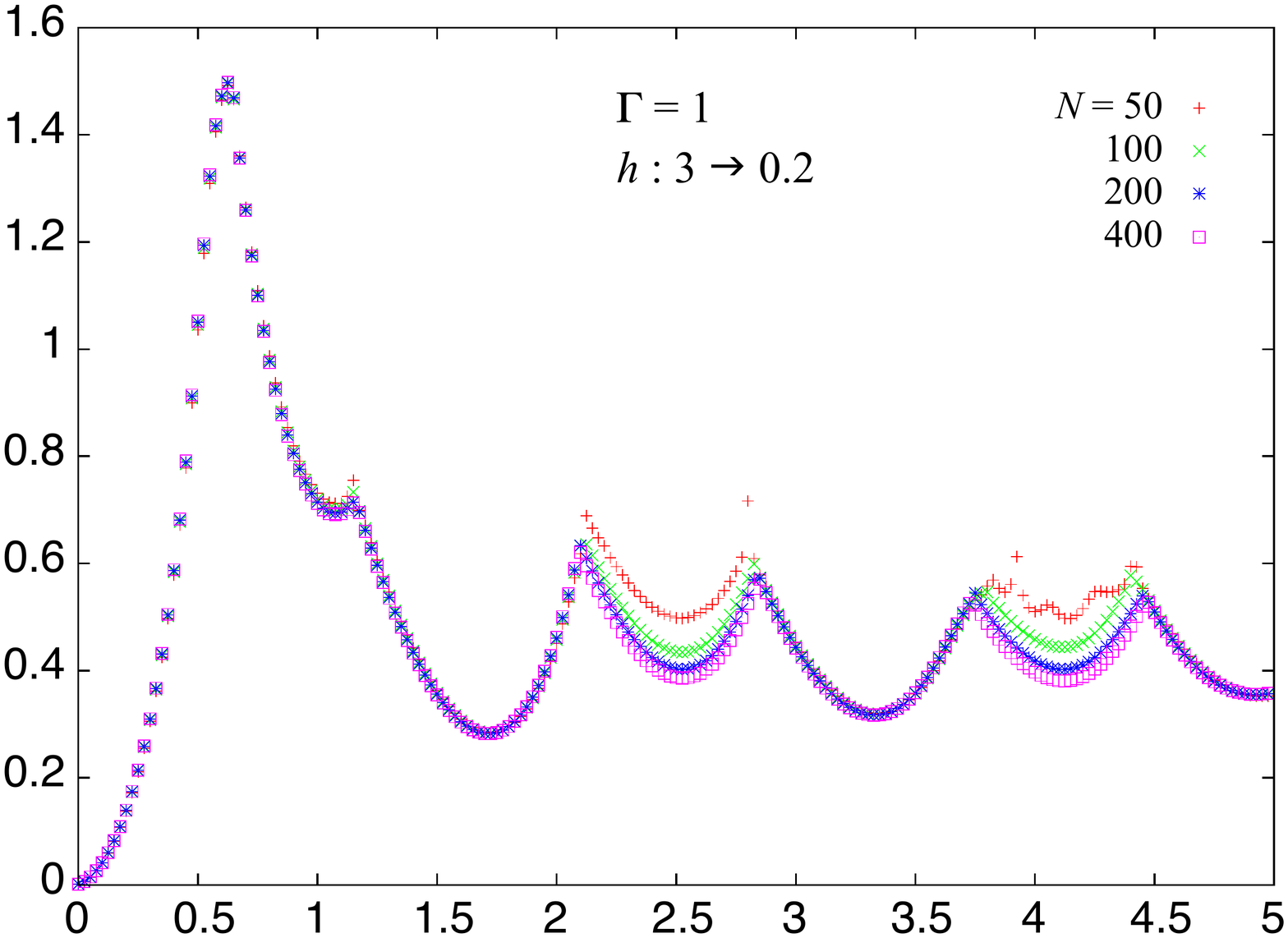}}
 %\label{fig:DPT_samephase_sud}
\subfigure[\ ]{
\includegraphics[width=4cm]{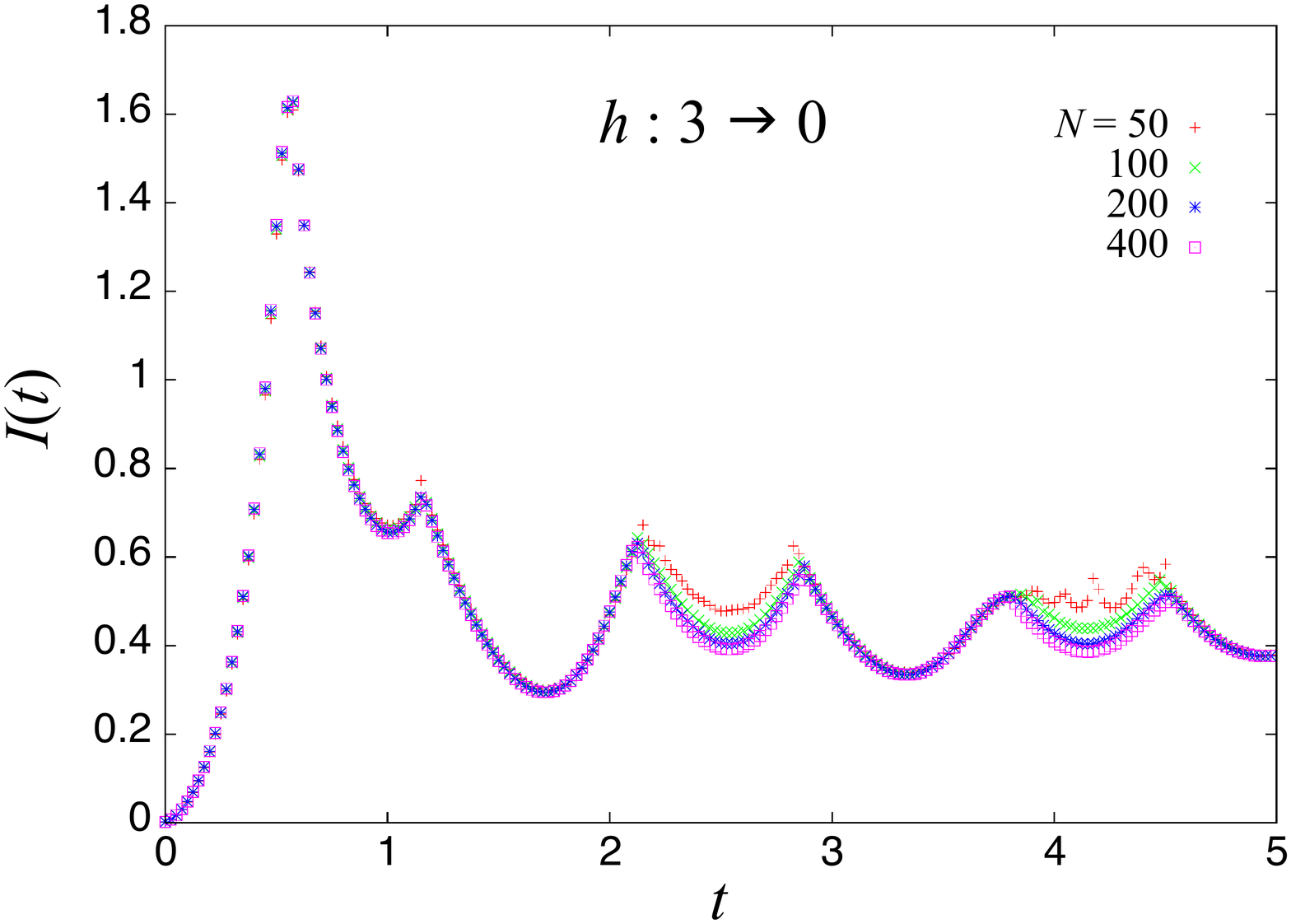}}
% \label{fig:DPT3_sud}
\subfigure[\ ]{
\includegraphics[width=4cm]{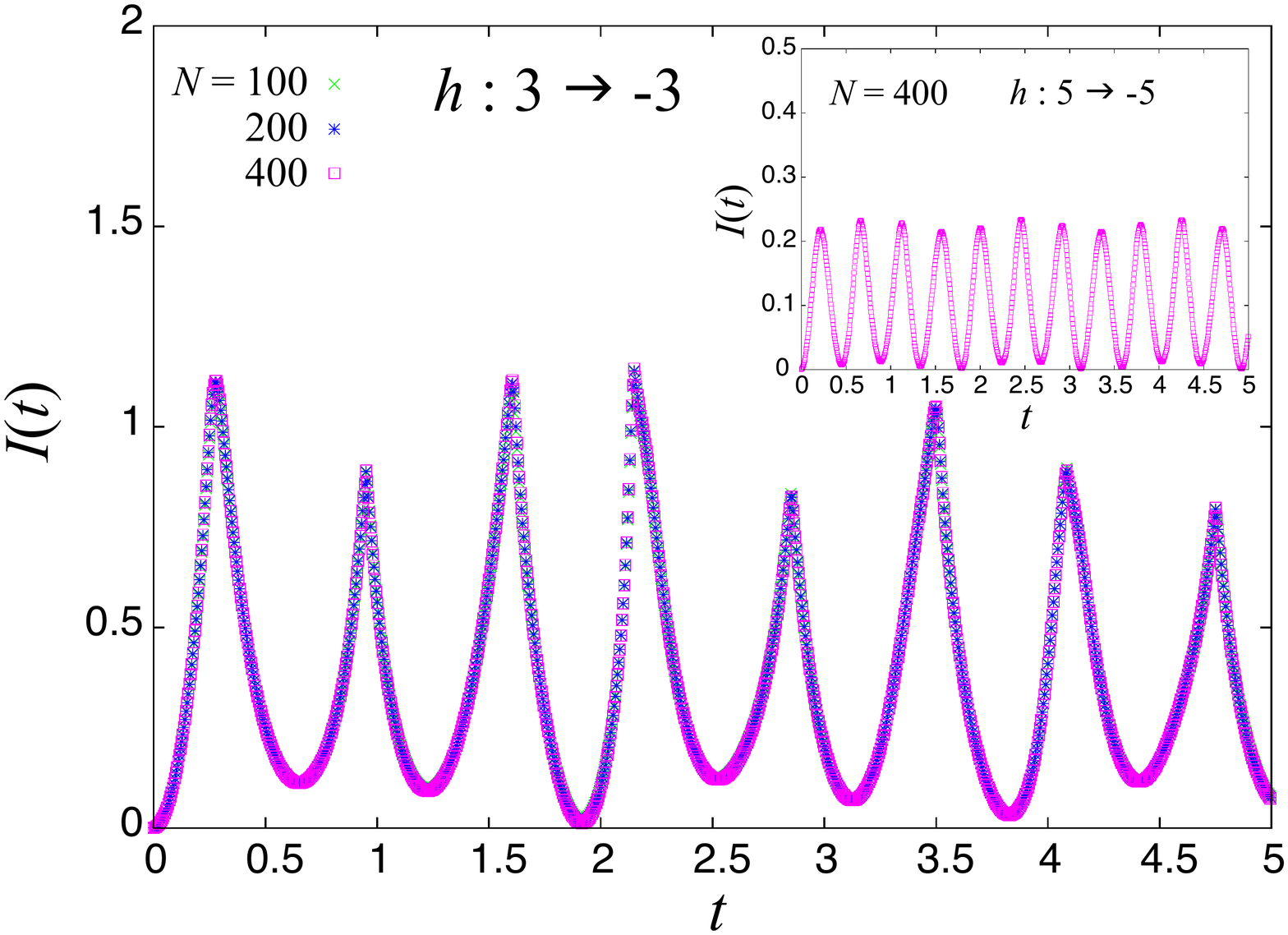}}
\caption{(Color online) Numerically obtained  $I(t)$ showing the absence and occurrence of the non-analyticities (DPTs) in different situations when $h$ is suddenly quenched. (a) No DPTs are observed for a sudden  quench of small amplitude of $h$ even if the system crosses the QCP in the process; (b) DPTs   occur when $h$ is quenched within the same phase; (c) DPTs also appear when $h$ is quenched from a large positive value to the QCP at $h=0$; (d) A regular  (but not periodic) occurrence of DPTs  is observed when $h$ is suddenly quenched from a large positive to a large negative value. The inset shows that the DPTs are rounded off when the quench amplitude is
even larger, leading to Rabi osscillations.
%\textcolor{red}{Sei, please include a new figure as 3c, where $h$ quenched from
%$h=1.5$ to the QCP at $h=0$, you have this data in the figure for slow
% quenching up to the QCP which we have removed.}
}
 \label{fig:DPT_sud}
 \end{figure}

Our results establish that for a sudden quench starting from the QCP {as well as a slow quench 
up to the QCP,  numerically obtained {residual energies
per spin}
exhibit   scaling relations which perfectly match earlier predictions. On the contrary}, there is  a series of interesting and unexpected results concerning the scenario of DPTs following these quenches  which are not reported before. Even though the model {is} non-integrable, we find prominent 
existence of DPTs when the longitudinal field is slowly ramped across the QCP.  {This is remarkable, given the fact that in
the FM case \ct{pollmann10} sharp non-analyticities are  present in $I(t)$ in the integrable case for $h=0$ when $\Gamma$ is ramped
across the QCP; on the contrary, those get smoothened out when the skewed field is quenched through the QCP at $\Gamma=1$ so that
the system is always non-integrable except at the QCP}.
On the other hand, for {sudden quenches the DPTs are found to occur whenever there is  a difference in the nature of the ground states of the initial and the  final Hamiltonians}
irrespective of the fact whether the system is quenched across a QCP or not.
%We observe non-analyticities for sudden quenching as 
%well as slow quenching of the longitudinal field when quenching starts and ends far away from the QCP whereas no such non-analyticities are observed 
%when quenching starts and ends close to QCP.  Also, DPTs are observed when quenching ends at QCP. The occurrence of DPTs also depends on the value of the 
%rate in the slow quenching case.

Let us first consider the situation when the field $h$ is ramped linearly to the
QCP ($h=0$)
%with a rate $\tau^{-1}$ 
as {$h = -t/\tau$} fixing
$\Gamma=1$. {
Denoting the final Hamiltonian $H_f$ {with ground state energy $E_f^0$,
and final wave function of the system (of length $N$) reached after the quench as $|\psi_f\rangle$}, 
the residual energy per spin is defined by
$\epsilon_{\rm res} = (\langle\psi_f |H_f |\psi_f\rangle - E_f^0)/{N}$.
%where $L$ denotes the system size.
%This quantity, as obtained numerically 
{Using
the t-DMRG calculations with
an open boundary condition, we find}}
$\epsilon_{\rm res} \sim \tau^{-4/3}$ (see main part of the Fig.~(\ref{fig:Eres})). This is in perfect agreement with the KZ scaling prediction, $\epsilon_{\rm res} \sim \tau^{-\nu(d+z)/(\nu z +1)}$ 
%(following a quench to the QCP)
 with $\nu=\nu_h =2$ and $z=1$ (see the
discussion in the SM). We now turn our attention to the sudden quench, in which 
the system is initially at the QCP and suddenly a small longitudinal
field $h$ is switched on; {in this case,
numerically we find}
  $\epsilon_{\rm res} \sim h^2$ (Inset, Fig.~(\ref{fig:Eres})). According to the prediction of the adiabatic perturbation theory \ct{degrandi10}, for such a sudden quench  of small magnitude starting from the QCP,
 $\epsilon_{\rm res}$ should scale as
$h^{\nu_h (d+z)}$, as long as the exponent does not exceed $2$; this is indeed true in the present case and as
a result the exponent saturates to  $2$.

Having established the scaling of $\epsilon_{\rm res}$ for both slow and sudden quenches, we now probe the scenario of possible
DPTs. 
%by tracking the temporal evolution of the final wave-function $\psi_ f$ reached following these quenches evolving 
%with the time-independent final Hamiltonian $H_{ f}$. 
The Loschmidt overlap at an instant $t$ ({where the initial time $t=0$ is
set immediately after the quenching is complete}) is  given by 
$G(t)= \langle \psi_f| \exp(-i H_f t) |\psi_f \rangle$ 
and consequently one defines  the rate function $I(t) = - \ln |G(t)|^2/N$, and investigates its temporal
evolution to probe the signature of possible DPTs (namely, the non-analyticities in $I(t)$)).  Results obtained for the slow
and sudden quenches obtained by using t-DMRG are presented  in Figs. \ref{fig:DPT2_tau} and \ref{fig:DPT_sud};
we below analyze the remarkable findings.

% pointing to the unexpected findings those we want to emphasize.
%Referring to the Fig. (\ref{fig:DPT_sud}), we find that there is no DPT or singularity in $I(t)$ in the rate function when $h$ is suddenly quenched from the initial value $h_i=+0.7$ to the final value $h_f=-0.7$. (To be precise, we have checked that DPT does not occurs when $h_i<1$  start showing up only when $h_i$ exceeds unity); this is surprising given the fact that the system indeed crosses the QCP at $h=0$ in the process. This is
%to be contrasted with the case when initial and final field is much larger than unity (e.g., $h_i=-3$ and $h_f=3$ as shown in Fig.~), when
%we find clear signature of DPT occurring nearly at regular intervals of time. Furthermore, remarkably one finds clear
%signature of DPTs for a sudden quenching with $h_i >>1 (=3$, as shown in Fig. ) which does not cross the QCP ($h_f=0.2$) or
%ends at the QCP ($h_f=0$).
%Surprising results also emerge when we explore what happens when the longitudinal field is quenched in a linear protocol characterized by an inverse rate
%$\tau$ starting from $h_i >>1$; DMRG results are presented in Fig.~(\ref{fig:DPT_tau}). One finds when quenched across the QCP, the magnitude
%of $I(t)$ decreases with $\tau$ while there is signature of DPTs even for higher values of $\tau$ (i.e., sufficiently slow
%quenching). On the contrary, when quenched to the QCP, the DPTs completely disappear when $\tau$ is larger. 

{We first analyze the slow-quenching of the model (\ref{eq_ham}) {with  $h \sim -t/\tau$,}
where $h$ is varied from a large positive to a large negative value.}  
 Referring to the Fig.~\ref{fig:DPT2_tau}, we find that  $I(t)$ shows non-analyticities
which appear at regular (and periodic) intervals  when $\tau \gg 1$ in contrast to the FM case \ct{pollmann10}.
%This is to be contrasted the situation presented in [\onlinecite{pollmann10}],
%where the presence of non-integrable term destroys the signature of DPT. 
To analyze this, we recall that for a sufficiently
slow driving the dynamics is always adiabatic except in the vicinity of the QCP ($h \ll 1$) where the relaxation time diverges. Remarkably,
the non-integrable Hamiltonian (\ref{eq_ham}) can be mapped to an effective integrable model {for $h\ll1$}, described the Hamiltonian:

\be
H_{\rm eff} = (1- bh^2) \sum_{i} \tau_i^z \tau_{i+1}^z - \sum_i \tau_i^x,
\label{eq:ham_eqv}
\ee
where $\tau_i$'s are Pauli spin matrices and $b$ is a constant which
{in our case can be chosen to be}  of the order of
unity; this mapping {to the model (\ref{eq:ham_eqv})} is
shown to
{exactly describe} the low-lying excitations of the Hamiltonian (\ref{eq_ham}) in the thermodynamic limit \ct{ovchinnikov03}.
Consequently so far as the slow quenching is concerned, when the dynamics is non-adiabatic only in the vicinity of a QCP, one can work  with the effective Hamiltonian (\ref{eq:ham_eqv})
which represents an AFM transverse Ising chain and is equivalent to a
FM transverse Ising chain by a simple
gauge transformation. Both the models are exactly solvable by Jordan-Wigner transformation. Focussing only at the QCP at 
$h=0$ and considering a slow ramp of $h$
 from a large {positive} value to a large {negative} value with the system initially in its ground state, one
can derive the final wave function by numerically integrating the corresponding Schr\"odinger equation; the rate function thus obtained indeed shows occurrences of the DPTs  thereby
qualitatively explaining the phenomena we observe here  (see the SM for details). 
%On the contrary, the FM version of the model \ct{pollmann10}
%can not be reduced to an integrable model close its QCP, and hence obviously one does not a priori expect  the occurrence of a DPT.

Interestingly,   the mapping to the effective Hamiltonian (\ref{eq:ham_eqv}) also enables us to explain the absence of
DPTs following a sudden quench of small amplitude across the QCP of the original model  as presented above in
Fig.~\ref{fig:DPT_sud}(a) because the interaction term in the equivalent Hamiltonian (\ref{eq:ham_eqv}) does not change sign which implies that this quenching
does \textit{not} take the system across the QCP of Hamiltonian (\ref{eq:ham_eqv}). 
%(one can further add that the system hits
%the the gapless point at $h=1$ and returns to the same phase without crossing it). 
This explains
the absence of DPT in this case though there is a crossing of QCP in the original model. Though the mapping to
the equivalent Hamiltonian is strictly valid for $ h\ll 1$, 
%our numerical finding shows that the DPT is present only when
%$h_i$ (or $h_f$) $\gg 1$ which can be explained using this Hamiltonian following a similar line of arguments; 
{in Fig.~\ref{fig:DPT_sud}(a), we show this argument can be extended} to explain the absence of DPTs when $h$ is quenched 
from $+0.7$ to $-0.7$ crossing the QCP at $h=0$.

Analyzing the original Hamiltonian (\ref{eq_ham}), we note that the ground state is paramagnetic with all spins polarized in 
the direction of $h$, when $h \gg 1$; on the contrary, it
{is} a quantum paramagnet with majority of spins orienting in
the direction of $\Gamma$ {when $h \ll 1$}.  The change in the nature of the ground state is reflected in DPT, irrespective of the 
fact whether the system crosses the QCP in the process of quenching. In Fig.~\ref{fig:DPT_sud}(b), we find a prominent presence of
DPTs when $h$ is quenched from {$3$} to $0.2$; here, even though the quenching does not take the original
Hamiltonian across a QCP, the nature of ground state changes. Similar DPTs are observed when quenched
to the QCP also (Fig.~\ref{fig:DPT_sud}(c)).  No such DPT is found to occur when the nature of the ground state is the
same (e.g., when  $h$ is changed from $3$ to $2$).
{Finally, when $h$ is suddenly quenched from $+3$ to
$-3$ across the QCP, one finds a regular ({but not periodic as  shown in Fig.~\ref{fig:DPT2_tau}) occurrence of DPTs (see Fig.~\ref{fig:DPT_sud}(c)). 
This is a generic feature of a sudden quench across the QCP as also observed
in the FM case \ct{karrasch13} (while the periodic pattern is only a characteristic
of the integrability of the underlying Hamiltonian).
%We note that a periodic pattern as seen in Fig.~\ref{fig:DPT2_tau} is
%not oberved in Fig.~\ref{fig:DPT_sud}(c); 
%it is the property observed only in the integrable limit. 
In this case, the initial and final ground states are nearly fully polarized states
%are paramagnetic with spins polarized 
%along $+h$ and $-h$ directions respectively, 
with their overlap being exponentially small with the system size;
this difference of the ground states results in observed DPTs.} 
When the quench amplitude is further increased (e.g., $h = +5$ to $-5$; see the inset of Fig.~\ref{fig:DPT_sud}(d)),
both the initial and final Hamiltonians essentially  reduce to an assembly of non-interacting spins;
in such situations  DPTs are rounded off leading to  Rabi oscillations between two fully polarized states.
% and replaced by the Rabi oscillation for a
%larger amplitude of the quench, for example, the quench from 
%. This is because 
%the Hamiltonians before and after the quench can be approximated by that of
%non-interacting spins. From this fact, it is inferred that
%the existence of a spin-spin interaction is crucial to DPT.
}
%(\textcolor{red}{Sei, 
%you are finding nearly periodic! Which means there is some underlying integrability. In our opinion for $h=3$ spins are
%almost non-interacting all the in direction of $h$; does it mean when you change the sign of $h$ there is
%an effective tunneling due to some effective $\Gamma$? We should mention this if we are absolutely sure; otherwise
%not!}) 

Finally, we summarize the results:
% of our study on the quenches and DPT of the model described by the Hamiltonian
%(\ref{eq_ham}):
 we have established that  $\epsilon_{\rm res}$ satisfies universal scaling 
relations for both sudden and slow quenches. {Furthermore, for the slow quenches, the model (\ref{eq_ham}) provides
a unique example where one can work with
an equivalent integrable model for $\tau \gg 1$}. This mapping enables us to explain the KZ
scaling and also a periodic occurrence of DPTs for a slow quenching across the QCP. This is remarkable {in the
sense}
that, to the best of our knowledge, the presence of DPTs following a slow quench of a non-integrable model has not been reported earlier; in the
FM situation, these non-analyticities get smoothened out 
%in the presence of an integrability breaking term 
\ct{pollmann10}.
Concerning the sudden quench, we also present some remarkable observations: in some cases, DPTs do not 
occur even when the system is quenched across the QCP;  but they may appear when the system is quenched
within the same phase (even to the QCP). {For very large amplitude quench of $h$ across $h=0$, DPTs get rounded off}.  {These
observations lead us to the conclusion that concerning the sudden quenches,
{it is} the change in the nature of {the} ground state  that is responsible for DPTs. }

{We would like to conclude with the note that the Hamiltonian
(\ref{eq_ham}) has been experimentally studied using Bose atoms in an optical-lattice \ct{simon11},
with $\Gamma \ll h$. 
% is tuned much smaller than the
%longitudinal field. 
The field $\Gamma$  of the equivalent spin chain is determined by the hopping amplitude $t$ of the Bose atoms and is
given by $2^{3/2}t$;  $\Gamma$ is necessarily kept small 
 to stablize the Mott state necessary for the realization of a spin system. On the other hand, a quantum Monte-Carlo
study \ct{rousseau06} shows that  in one dimension
% its critical value is large enough to have 
it should be possible to achieve a field $\Gamma \approx 1$.
Therefore it should be possible to verify some of  the situations
of the present study in experimental systems.}

%4. Possible scenario of dynamical phase transitions\\

%Acknowledgement
We acknowledge Jun-ichi Inoue  for discussions. Shraddha Sharma acknowledges
CSIR, India and AD acknowledges DST India for financial support.
SS thanks MEXT, Japan for support through grant No. 26400402.

%%%%%%%%%%%%%%%%%%%%%%%%%%%%%%%%%%%%%%%%%%%%%%%%%%%%%%%%%%%%%%%
\vspace{-\baselineskip}

%\end{document}

\clearpage
%\widetext
%\begin{center}
%\textbf{\large Supplementary Material on ``Quenches and dynamical phase transition in a non-integrable quantum Ising model''}\\
%\vspace{0.5cm}
%{Sei Suzuki$^1$, Shraddha Sharma$^2$ and Amit Dutta$^2$}\\
%$^1$ Department of Liberal Arts, Saitama Medical
%University, Moroyama, Saitama 350-0495, Japan\\
%$^2$Department of Physics, Indian Institute of
%Technology, Kanpur 208 016, India
%\end{center}
%
%
%%\setcounter{equation}{0}
%%\setcounter{figure}{0}
%%\setcounter{table}{0}
%%\setcounter{page}{1}
%%\makeatletter
%%\renewcommand{\theequation}{S\arabic{equation}}
%%\renewcommand{\thefigure}{S\arabic{figure}}
%%\renewcommand{\bibnumfmt}[1]{[S#1]}
%%\renewcommand{\citenumfont}[1]{S#1}
%
%
%
%In this supplementary material, we shall illustrate how to derive the Kibble-Zurek scaling relation for the residual
%energy of the original Hamiltonin (\ref{eq_ham}) and also show how the behavior of the Fisher zeros in the
%complex time plane dictate the occurrence of DPTs for such a slow quench.
%
%\setcounter{equation}{0}
%\setcounter{figure}{0}
%\setcounter{table}{0}
%\setcounter{page}{1}
%\makeatletter
%\renewcommand{\theequation}{S\arabic{equation}}
%\renewcommand{\thefigure}{S\arabic{figure}}
%\renewcommand{\thesection}{S\arabic{section}}
%\renewcommand{\bibnumfmt}[1]{[S#1]}
%\renewcommand{\citenumfont}[1]{S#1}
%
%%%%%%%%%%%%%%%%%%%%%%%%%%%%%%%%%%%%%%%%%%%%%%%%%%%%%%%%%%%%%%%
%\end{document}
%%%%%%%%%%%%%%%%%%%%%%%%%%%%%%%%%%%%%%%%%%%%%%%%%%%%%%%%%%%%%%%

\widetext
\begin{center}
\textbf{\large Supplementary Material on ``Quenches and dynamical phase transition in a non-integrable quantum Ising model''}\\
\vspace{0.5cm}
{Shraddha Sharma$^{1}$, Sei Suzuki$^{2}$  and Amit Dutta$^{1}$ }\\
\vspace{0.2cm}
{$^1$}{\it Indian Institute of Technology Kanpur, Kanpur 208 016, India} \\
{$^2$}{\it Department of Liberal Arts, Saitama Medical
University, Moroyama, Saitama 350-0495, Japan} \\
\end{center}

\setcounter{equation}{0}
\setcounter{figure}{0}
\setcounter{table}{0}
\setcounter{page}{1}
\makeatletter
\renewcommand{\theequation}{S\arabic{equation}}
\renewcommand{\thefigure}{S\arabic{figure}}
\renewcommand{\bibnumfmt}[1]{[S#1]}
\renewcommand{\@cite}[1]{[S#1]}

\begin{figure}[h]
\begin{center}
 \includegraphics[width=3.0in]{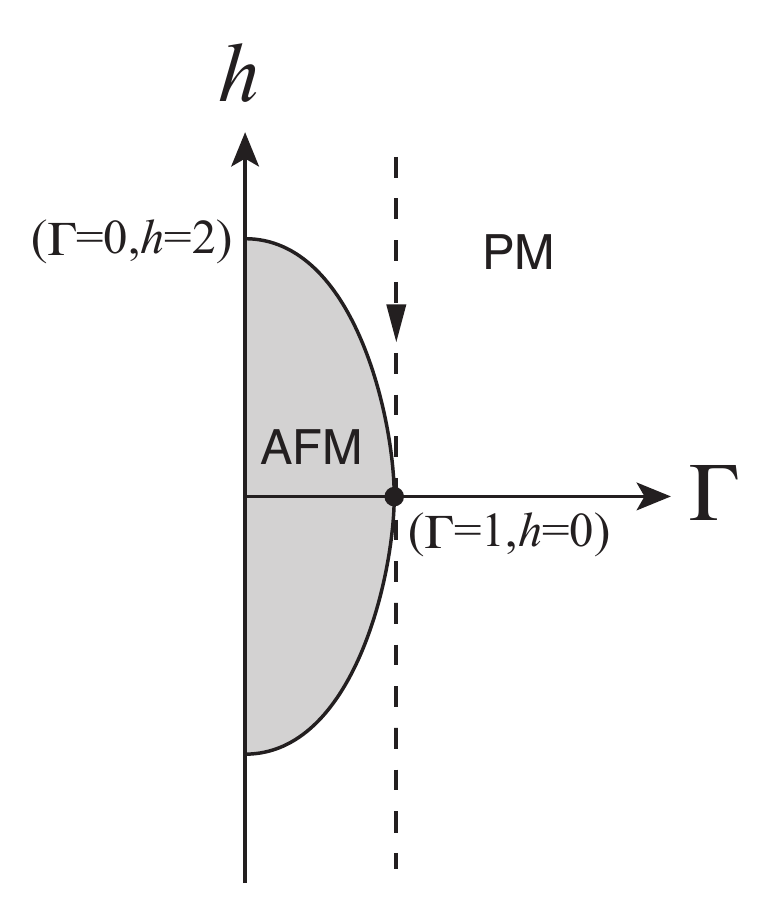}
\end{center}
\caption{The {schematic} phase diagram of the model given in
 Eq.~(\ref{suppl:eq_ham}); the solid line extending from
 $(\Gamma=0,h=2)$ {to $(\Gamma=0,h=-2)$ through
 $(\Gamma=1,h=0)$}
separates the antiferromagnetic (AFM) phase from the paramagnetic (PM)
 phase. {The points $(\Gamma=0,h=\pm 2)$}
denotes the first order transition while   $(\Gamma=1,h=0)$  corresponds
 to the integrable quantum critical point. Throughout
 {this paper,
$\Gamma$ is set equal to $1$ and  $h$ is quenched along the dashed line shown with an arrow.}
\label{fig:sup1}
}
\end{figure}

{\small In this supplementary material, we shall illustrate how to derive the Kibble-Zurek scaling relation of the residual
energy of the original Hamiltonian
\be
H= \sum_i \si_i^z \si_{i+1}^z - \Gamma \sum_i \si_i^x - h \sum_i \si_i^z,
\label{suppl:eq_ham}
\ee
when the longitudinal field $h$ is quenched as $-t/\tau$,
and also show how the behavior of the Fisher zeros in the
complex time plane dictates the occurrence of DPTs for such a slow quench. The transverse field $\Gamma$ is always
set to unity so that the system is at the integrable quantum critical point (QCP) when $h=0$. The phase diagram of
the model and the quenching path are shown in Fig.~\ref{fig:sup1}.
}

%Finally, In Sec.~\ref{exp}, we given an experimental realization of the generation of complex hopping term. 

%%%%%%%%%% Merge with supplemental materials %%%%%%%%%%
%%%%%%%%%% Prefix a "S" to all equations, figures, tables and reset the counter %%%%%%%%%%

\section{The Kibble-Zurek Scaling}
\label{supple_kibble_zurek}

As emphasized in the main text, so far as the slow quenching of the longitudinal field $h$ is concerned (especially,
for large $\tau$), one can equivalently work with the effective integrable Hamiltonian given by

\ba
H_{\rm eff} &=& (1- b h^2) \sum_{i} \tau_i^z \tau_{i+1}^z - \sum_i \tau_i^x,
\label{eq:ham_eqv_sub}
\ea
where $b$ is a constant \cite{suppl:ovchinnikov03} which  is  inessential in the argument below,
and hence set equal to unity hereafter.
Using a gauge transformation (which flips the spins of alternate sites) and a duality transformation \ct{suppl:kogut79S}, the Hamiltonian
in Eq.~(\ref{eq:ham_eqv_sub}) can be mapped to an equivalent dual Hamiltonian with a nearest
neighbor {ferromagnetic (FM)} interactions:

\ba
{\tilde H}_{\rm eff} &=& - \sum_{i} {\tilde \tau}_i^z {\tilde \tau}_{i+1}^z -  (1-h^2) \sum_i {\tilde \tau}_i^x \non\\
&=& - \sum_{i} {\tilde \tau}_i^z {\tilde \tau}_{i+1}^z -   \sum_i {\tilde \tau}_i^x + h^2 \sum_i {\tilde \tau}_i^x\label{eq:ham_dual}
\ea
which is a FM transverse Ising Hamiltonian in an effective transverse field $\Gamma_{\rm eff} = 1- h^2$.
We note that  ${\tilde H}_{\rm eff}$ with $h=0$ represents a critical
Hamiltonian. Using {the} Fourier transformation
followed {by the} Jordan-Wigner transformation, the model can be
reduced to a two-level problem in the basis $|0\rangle$ (no fermion state ) and $|k,-k\rangle$ (a state with a pair of fermions with
quasi-momenta $k$ and $-k$, respectively) \ct{suppl:Sachdev,suppl:suzuki13}; the reduced $2\times 2$ Hamiltonian is then
given by
\begin{equation}
 H_{k} (h)= 
2 \left(
 \begin{array}{cc}
 (1-h^2) -\cos k  & -i \sin k  \\
i \sin k  &   -(1-h^2) +\cos k    \\
 \end{array}
 \right).
 \label{ham_2by2_full}
 \end{equation}

Analyzing the spectrum , $\epsilon_k=  2 \sqrt{ \{(1-h^2) -\cos k\}^2 + \sin^2 k}$, it is  straightforward to show 
that the model (\ref{eq:ham_dual}) has three QCPs; the energy gap ($2 \epsilon_k$)
 vanishes at  critical points at $h=0$ and $h=\pm \sqrt{2}$, with the
corresponding critical wave vector (for which the energy gap vanishes)
$k_c=0$ and $\pi$, respectively. We
are however interested in the transition at $h=0$ which is the only relevant QCP to the context of  the original Hamiltonian (\ref{suppl:eq_ham}).
To focus on the critical point at $h=0$, we expand the Hamiltonian (\ref{ham_2by2_full}) in the vicinity
of $k=0$ to arrive at the Hamiltonian

\begin{equation}
 H_{k} (h)= 
 2\left(
 \begin{array}{cc}
 -h^2 + \frac{k^2}{2} & -ik  \\
ik  &  h^2 -\frac{k^2}{2}   \\
 \end{array}
 \right),
 \label{ham_2by2}
 \end{equation} 
 which shows only one quantum critical point at $h=0$.
 Analyzing the simplified form of the spectrum $\epsilon_k = \sqrt{(h^2 -k^2/2)^2+ k^2}$,  one immediately finds for $h=0$, the
 gap ($\Delta E_k =2\epsilon_k$) $\sim k$, yielding $z=1$ and for $k=0$, gap scales as $h^2$ yielding $\nu z=2$, and hence  $\nu=2$ (referred
 to as $\nu_h$ in the main text).

Let us now point out  that the  quenching $h= -t/\tau$, with $t$ going from $-\infty$ to $0$, in the original Hamiltonian is equivalent to driving the reduced
Hamiltonian (\ref{ham_2by2}) from $h \to \infty$ to the QCP at $h=0$ by  a non-linear protocol $(t/\tau)^2$; in both the cases
the system is initially prepared in its ground state. Even though the non-adiabatic transition probability for the
mode $k$ ($p_k$) can not be calculated directly using the Landau-Zener formula for such a non-linear protocol, one can make appropriate rescaling in  the corresponding Schr\"odinger equations \ct{suppl:mondal08,suppl:dutta15}
to  argue that it  would be a function of the dimensional combination of
$k^2 \tau^{4/3}$, i.e., $p_k = {\cal F}(k^2 \tau^{4/3})$ where ${\cal F}$ is an unknown scaling function. Since the gapless QCP is characterized by gapless  excitations 
$k$, the scaling of the residual energy can be obtained as $\epsilon_{\rm res} \sim \int dk k {\cal F}(k^2 \tau^{4/3}) \sim \tau^{-4/3}$;
this matches perfectly with the KZ prediction with $d=z=1$ and $\nu=2$ and the numerical result presented  in the main text.

\section{Slow quenching and non-analyticities in the rate function}

\label{sec_fisher_zero}

We shall now calculate the nature of  the Fisher zeros of the effective partition function \ct{suppl:heyl13} obtained from the Loschmidt overlap (LO) when the parameter $h$ of the Hamiltonian (\ref{eq:ham_dual}) is quenched from a large positive to a large negative value following the protocol $h=-t/\tau$; this is equivalent to the slow quenching of the longitudinal field  $h$ {in} the original Hamiltonian (\ref{suppl:eq_ham}). But there is a subtle difference that needs to be emphasized: the field $h$ {contributes a quadratic} $h^2$ term to the transverse field of the equivalent
model  (\ref{eq:ham_dual}), thus as $h$ is linearly changed from a large positive value to the negative value in model
(\ref{suppl:eq_ham}), the parameter $h^2$ changes from
a positive initial value to zero (i,e, the QCP) and returns to the original initial value at the final time; this in a sense is a reverse quenching of the transverse field of the Hamiltonian (\ref{eq:ham_dual})  as studied in \ct{suppl:divakaran09}  in a non-linear
fashion.

\begin{figure}
%\begin{center}
 \includegraphics[width=3.2in]{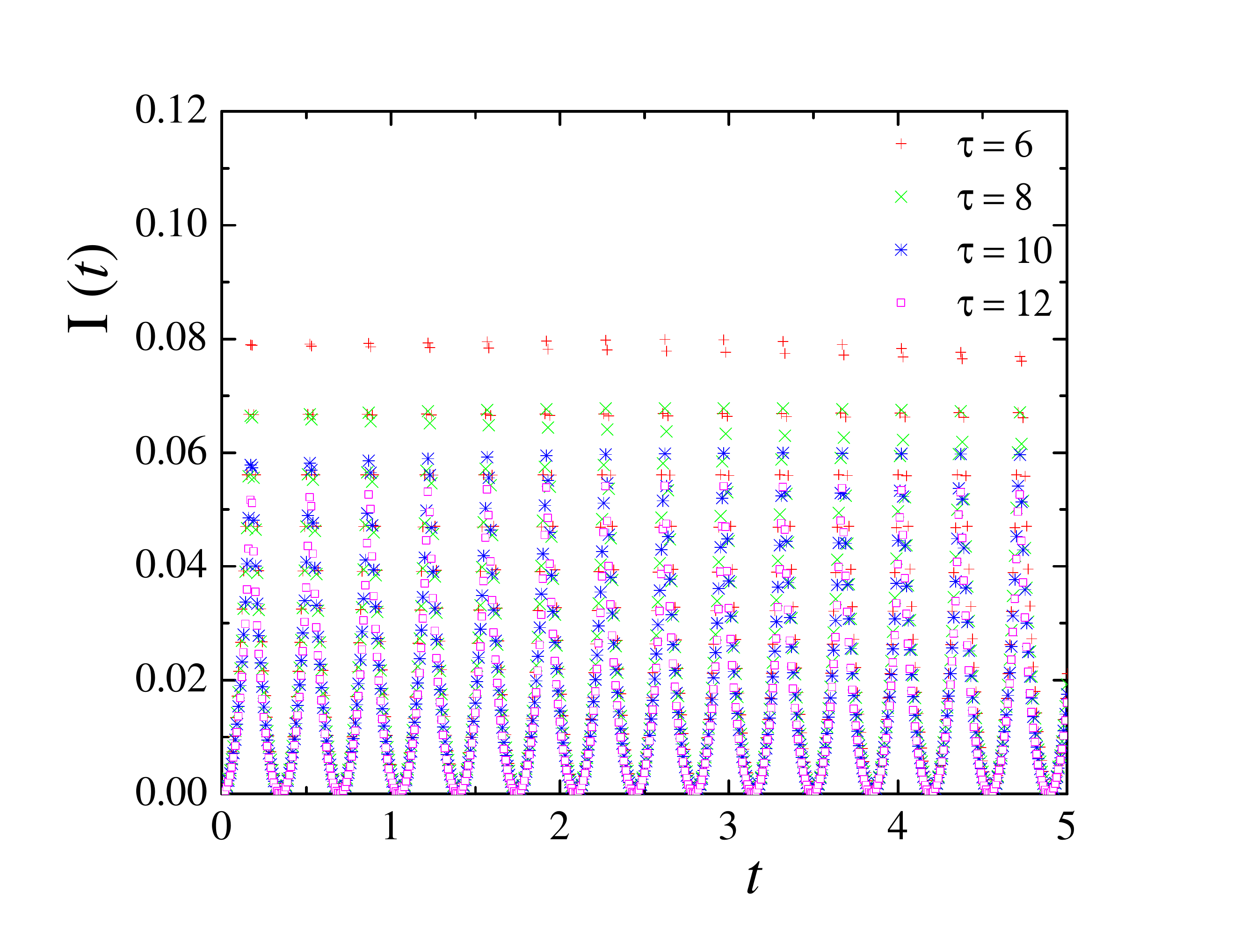}
% \end{center}
\caption{The rate function for quenching $h=3$ to $h=-3$ shows sharp
 non-analyticities at periodic intervals in time with system 
 size $N=400$ and several $\tau$'s. }
     \label{fig:DPT_integrable}
\end{figure}

To calculate the  Loschmidt overlap  of a system of length $N$ defined by $f(z) = - \ln \langle \psi_f |\exp(-H_f z)| \psi_f \rangle/N$, where $z$ is the
complex time, $H_f$ is the final Hamiltonian and $|\psi_f \rangle$ the state reached following the quantum quench, we focus
on the reduced $2 \times 2$ Hamiltonian (\ref{ham_2by2}). Summing over the contributions from all the momenta mode, a few lines
of algebra  leads us to the expression \ct{suppl:sharma15}
\ba
f(z) =  - \int_0^{\pi} \frac{dk}{2\pi} \ln \left( (1-p_k) + p_k \exp(-2 \epsilon_k^f z) \right)\ea
where $p_k$ is the non-adiabatic transition probability for the mode $k$. The  zeros of the ``effective'' partition  function (where
$f(z)$ is non-analytic) are given by:

\be
z_n(k) = \frac 1 {2 \epsilon_k^f}  \left( \ln (\frac { p_k}{1-p_k}) + i \pi(2n+1)\right),
\label{eq_fisher_zero}
\ee
where $n=0,\pm 1, \pm 2, \cdots $.
{For a non-linear reverse quenching protocol,} the expression for $p_k$ can not be exactly determined using the LZ
formula (though an exact form can be obtained for the linear case \ct{suppl:divakaran09}). However, it can be
argued $p_k = {\cal G}((k-k_0)^2 \tau^{4/3}$), where $k_{0}$ is the wave vector for which $p_k$ is maximum which
shifts to $k=0$ for large $\tau$ and ${\cal G}$ is an unknown function. We find from Eq.~(\ref{eq_fisher_zero}) Fisher zeros  cross the imaginary axis for a particular value of $k_*$ for which $p_{k_*} =1/2$ \ct{suppl:sharma15,suppl:pollmann10} and 
the rate function shows sharp non-analyticities at   $ t_n^* = {\pi (n+\frac 1 {2})/\epsilon_{k_*}^f}$.
For the present case, to calculate the Fisher zeros and
especially the rate functions $I(t)$  we shall use the form of the Hamiltonian  near $k=0$ given in Eq.~(\ref{ham_2by2})
(to avoid the influence of the QCP at $h=\sqrt {2}$),
%However, to avoid the influence of the QCPs at $h=\pm \sqrt{2}$, we shall restrict our initial and final field with $|h| \ll \sqrt{2}$
%In the following, we present results 
when $h$ is quenched from $+3$ to $-3$.  Numerically integrating
the Schr\"odinger equation describing the dynamics of {the Hamiltonian} (\ref{ham_2by2})  with the initial condition that the system
is in the ground state of the initial Hamiltonian, we obtain the value of $p_k$ which is then substituted in the expression of
 the rate function:  
\be
I(t)= - \int_{0}^{\pi} \frac{dk}{2\pi} \log\left(1 + 4 p_k (p_k-1) \sin^2 \epsilon_k^f t \right).
\ee
 As shown in 
Fig. \ref{fig:DPT_integrable}, this qualitatively explains 
the periodic occurrence of DPTs presented in
Fig.~\ref{fig:DPT2_tau} in the main text.

% In Fig.~(\ref{fig:DPT_integrable}a), we  plot Eq.(\ref{eq_fisher_zero}) in the real and imaginary time plane and show that indeed they cross the imaginary time axis for a particular value of $k$ and the rate function $I(t)$ show non-analyticties
%at corresponding values of real time as shown in Fig.~(\ref{fig:DPT_integrable}b).

%\begin{figure}[h!]
%\centering
%\subfigure[\  Fisher zeros crossing the imaginary axis for quenching of the longitudinal field from $-1 \to 1$ in the equivalent  model
%(\ref{ham_2by2_full}) for $\tau=6$ and $N=400$. Since the final value of $h=1$, and hence $\epsilon_k^f =1$, the imaginary
%part of $z_n(k)$ is independent of $k$ is given by $\pi(n+1/2)$.]{
%\includegraphics[width=6cm, 0 0 764 586]{figures/DPT_tau_6.pdf}}
%% \label{fig:DPT07_sud}
%\subfigure[\ The rate function shows non-analyticities at time instants $t_n$ with different values of $\tau$ as shown for $N=400$.]{
%\includegraphics[width=6cm, 0 0 764 586]{figures/Rate_Full.pdf}}
% %\label{fig:DPT_samephase_sud}
%%\subfigure[\ DPTs on the other hand are observed when the longitudinal field is suddenly quenched from a large negative
%%to a large positive value.]{
%%\includegraphics[width=4cm]{figures/AFG10h30--30RateF.pdf}}
%%% \label{fig:DPT3_sud}
%\caption{ The Fisher zeros and the rate functions obtained  for the model (\ref{ham_2by2_full}) are as shown above.}
% \label{fig:DPT_integrable}
% \end{figure}

\end{document}